\documentclass[prx,reprint,amsmath,amssymb,aps,floatfix]{revtex4-2}
\usepackage{url}
\usepackage{hyperref}
\usepackage{graphicx}
\usepackage{dcolumn}
\usepackage{xcolor}
\usepackage{soul}
\usepackage{bm}
\begin{document}
\preprint{APS/123-QED}

\title{Correlated frequency noise in a multimode acoustic resonator}

\author{Nuttamas Tubsrinuan}
\email{nuttamas.tubsrinuan.20@ucl.ac.uk}
\affiliation{Department of Microtechnology and Nanoscience MC2, Chalmers University of Technology, Göteborg, Sweden}
\affiliation{London Centre for Nanotechnology, University College London, Gordon Street, London, WC1H 0AH, UK}

\author{Jared H. Cole}
\affiliation{Chemical and Quantum Physics, School of Science, RMIT University, Melbourne, VIC 3001, Australia}

\author{Per Delsing}
\affiliation{Department of Microtechnology and Nanoscience MC2, Chalmers University of Technology, SE-41296 Göteborg, Sweden}

\author{Gustav Andersson}
\email{gandersson@uchicago.edu}
\affiliation{Department of Microtechnology and Nanoscience MC2, Chalmers University of Technology, Göteborg, Sweden}
\affiliation{Pritzker School of Molecular Engineering, University of Chicago, Chicago, Illinois 60637, USA}

\date{\today}

\begin{abstract}
Frequency instabilities are a major source of errors in quantum devices. This study investigates frequency fluctuations in a surface acoustic wave (SAW) resonator, where reflection coefficients of 14 SAW modes are measured simultaneously for more than seven hours. We report two distinct noise characteristics. Multimode frequency noise caused by interactions with two-level system (TLS) defects shows significant degrees of correlations that diminish with increased detuning. This finding agrees with the current understanding of the parasitic TLS behavior as one of the dominant noise sources in quantum devices. In addition to the TLS-induced noise, we observe strong anomalous frequency fluctuations with slow, anti-correlated dynamics. These noise bursts resemble signatures of cosmic radiation observed in superconducting quantum systems.
\end{abstract}
\maketitle

\par Decoherence in quantum devices leads to quantum information loss and limits their performance \cite{Simmonds2004, Burnett2019}. A number of decoherence channels have been identified, including various types of parasitic two-level systems (TLSs) \cite{Muller2015, Martinis2005} and quasiparticles \cite{Wilen2021}. Quasiparticle generation can be induced by stray infrared light \cite{Barends2011} and cosmic radiation \cite{Vepsalainen2020}, while TLS defects inducing loss and noise reside in the circuit \cite{Burnett2014, Muller2019}. Over long timescales, these noise sources impact the stability of device parameters \cite{Muller2015, Klimov2018, Burnett2019}, necessitating frequent calibration or additional mitigation measures \cite{Matityahu2021, You2022}. In the past decades, significant enhancement in the coherence of superconducting devices has been achieved by improved microwave engineering \cite{Chiaro2016}, circuit design \cite{Koch2007},  fabrication methods \cite{Cicak2010}, and materials \cite{Chang2013}. In spite of these developments, decoherence remains a significant issue. A better understanding of noise processes is important in order to mitigate decoherence mechanisms and advance the development of solid state quantum processors.
\par In this work, we use a multifrequency signal (a frequency comb) to simultaneously probe frequency fluctuations in 14 modes of a surface acoustic wave (SAW) resonator at cryogenic temperatures. Tracking the reflection coefficient of the SAW resonator for more than 7 hours, we observe that there are two dominant noise processes taking place in the device. We further investigate the noise correlations and their characteristics. 
\par Two-level system defects residing at dielectric surfaces and interfaces of superconducting circuits have been considered a central problem causing frequency fluctuations and significant energy dissipation \cite{Martinis2005, Burnett2014, Martinis2005, Lisenfeld2019}. Despite many attempts in the past decades \cite{Muller2019}, TLS-induced loss and noise remain persistent issues in superconducting quantum processors. Although the microscopic nature of TLSs is not well understood, the standard tunneling model (STM) \cite{Phillips1972, Anderson1972, Muller2019} has been successfully used as a framework for understanding TLS interaction with quantum circuits. The STM predicts power-dependent losses and temperature-dependent resonance frequency shifts due to the direct coupling between quantum devices and coherent TLSs, which have transition energies larger than $k_BT$. In addition to coherent TLSs there are thermal fluctuators, two-level systems with sufficiently low transition energies to be thermally activated at millikelvin temperatures. Direct interaction between TLSs is a significant effect not accounted for by the STM \cite{Burnett2014,Faoro2015,Lisenfeld2015}. When coherent TLSs couple with surrounding thermal fluctuators, they are subject to a continuous time-dependent energy shift, known as spectral diffusion \cite{Black1977}. While undergoing spectral diffusion, they can couple dispersively to quantum devices, resulting in frequency drift that translates into phase noise \cite{Gao2007, Niepce2021}.
\par TLSs couple to both electromagnetic \cite{Martinis2005} and strain fields \cite{Grigorij2012}. Their behaviors have been studied using superconducting qubits \cite{Martinis2005, Lisenfeld2015}, microwave resonators \cite{Gao2008, Pappas2011}, and mechanical resonators \cite{Anghel2007, Andersson2021, Cleland2023}. A study of a surface acoustic wave (SAW) device at cryogenic temperature reveals that energy losses are caused by the coupling to parasitic TLSs \cite{Manenti2016}. Further two-tone spectroscopy of TLSs using a SAW resonator shows that TLSs dominate the losses \cite{Andersson2021}, suggesting the potential of SAW resonators for investigating TLSs. Due to the slow propagation, the SAW wavelength is smaller than the electromagnetic wavelength by five orders of magnitude at the same frequency. This characteristic allows SAW resonant structures to accommodate many resonant frequencies in a compact design.
\par Here, we exploit multiple resonance modes of the SAW resonator to probe frequency noise caused by TLSs at many different frequencies simultaneously. According to the Jaynes-Cummings model \cite{Jaynes1963}, the resonance frequency of a resonator $f$, which is dispersively coupled to an individual TLS, is given by 
\begin{equation}
    f = f_r + \frac{g^2}{f_r-f_{TLS}}\hat{\sigma}_z,
\label{Eqn_JC}
\end{equation}
where $\hat{\sigma}_z$ is the Pauli spin operator acting on the TLS, and $g$ represents the TLS-resonator coupling strength. The resonance frequency of the bare resonator and the TLS are denoted by $f_r$ and $f_{TLS}$, respectively. In most realistic cases, the resonator interacts with a TLS ensemble with a frequency distribution spanning the resonator bandwidth. The effective frequency shift is the sum of the dispersive shifts contributed by all the individual TLSs. Due to their small wavelengths and high wavenumbers, SAW modes are close to each other in frequency and have a large overlap in real space. Thus, all modes interact with the same physical TLS ensemble, however, with slightly different detuning. Therefore, the multimode frequency noise in a SAW resonator is expected to be correlated. Additionally, the degree of correlation should depend on the detuning, as suggested by Eq.~\ref{Eqn_JC}.

In superconducting devices, cosmic rays and other sources of radiation can generate high-energy phonons \cite{Grunhaupt2018,Swenson2010, Moore2012}. These phonons break Cooper pairs, leading to an increased kinetic inductance of the superconducting material and associated resonance frequency shifts \cite{Catelani2011}. A recent study on superconducting granular aluminum resonators shows that ionizing radiation due to high-energy cosmic rays create phonons that in turn generate quasiparticle bursts, resulting in resonance frequency shifts and reduced quality factors \cite{Cardani2021}. This effect has also been observed in transmon qubits with shortened relaxation and dephasing times as a result \cite{Vepsalainen2020,deGraaf2020,Mcewen2021}. Direct measurements of rare events which were attributed to cosmic rays have also been observed in gravitational wave experiments \cite{Astone2001, Astone2008, Goryachev2021}. The frequency noise caused by ionizing radiation can be suppressed by the use of a phonon absorber \cite{Karatsu2019}, radiation shielding, and conducting experiments at a deep-underground facility \cite{Cardani2021}.
\par
\begin{figure}
\includegraphics[width = \linewidth] {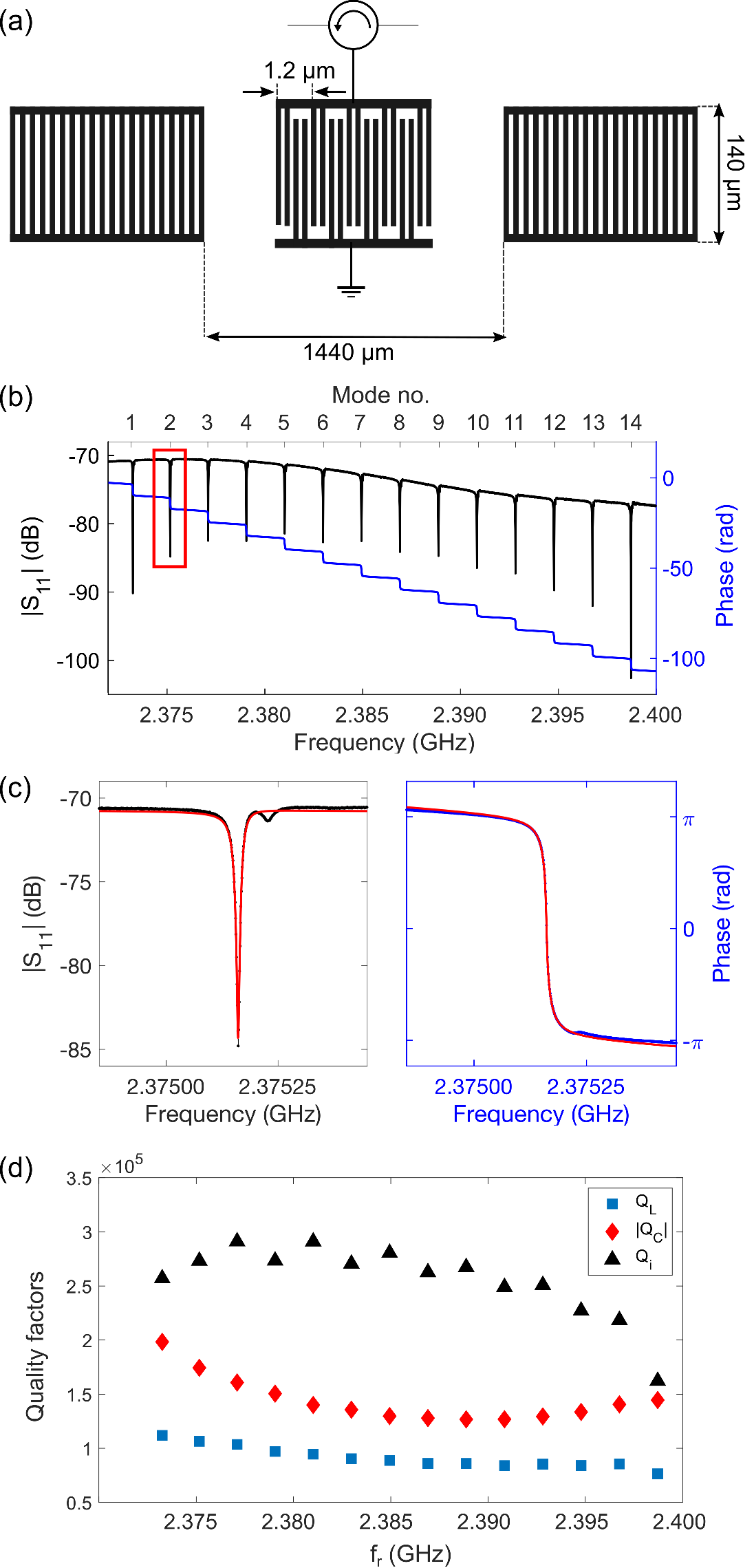}
\caption{(a) Dimensions of the SAW resonator used in this study (not to scale). The mirrors and the IDT are fabricated from aluminum patterned on a GaAs substrate. The shorted Bragg mirrors have 800 fingers each.  The IDT has a split-finger design with 50 periods of $1.2\ \mu$m. (b) The reflection measurement of the SAW resonator, showing the 14 resonance modes lying between the frequency of 2.37 and 2.40 GHz. The indices represent mode numbers arranged from the lowest to the highest frequency of the resonator stopband. (c) The fitted magnitude and phase of resonance mode 2 highlighted by the red box in the top figure. The quality factors are extracted using Eq.~\ref{S11eqn}. (d) SAW mode quality factors. The coupling coefficient $Q_C$ ranges from $1.2 \times 10^5$ to $2.0 \times 10^5$ and the internal quality factor $Q_i$ varies from $2.0 \times 10^5$ to $2.5 \times 10^5$.}
\label{fig_SAWprofile}
\end{figure}

 Our SAW resonator, fabricated on a piezoelectric GaAs substrate, has two Bragg mirrors, located 1440 $\mu$m apart. Each mirror has 800 fingers. An interdigital transducer (IDT) is placed at the center of the cavity, serving as both input and output port. The IDT has a split-finger design with 50 periods of 1.2 $\mu$m, as illustrated in Fig. ~\ref{fig_SAWprofile}(a) \cite{Andersson2021}. The device supports 14 resonance modes within the frequency range from 2.37 GHz to 2.40 GHz. The free spectral range between each mode is approximately 2 MHz.
The characteristics of the SAW resonator are shown in Fig.~\ref{fig_SAWprofile}(b)-(c). The resonance linewidths are between 18 and 32 kHz. We extract the resonator parameters based on the expression for the reflection coefficient \cite{Probst2015}:
\begin{equation}
S_{11}(f) = -ae^{-i(\pi-\theta)}\left[1-\frac{2(Q_L/Q_C)e^{i\phi_0}}{1+2iQ_L(\frac{f-f_r}{f_r})}\right],
\label{S11eqn}
\end{equation}
where $f$ is the probe frequency and the parameter $a$ is an amplitude scaling coefficient. A phase shift due to cable delay is denoted by $\theta$. The loaded quality factor and the external quality factor are represented by $Q_L$ and $Q_C$, respectively. An impedance mismatch is accounted for by the parameter $\phi_0$. We drive the SAW resonator continuously so that the estimated number of phonons is approximately 1000. The probe frequencies are defined from the resonance frequencies extracted from the resonator profile in Fig.~\ref{fig_SAWprofile}(b) using Eq.~\ref{S11eqn}, such that the phase response is maximally sensitive to frequency shifts.
\par For this device, the quality factors are slightly different for each frequency, as shown in Fig.~\ref{fig_SAWprofile}(d). We measure the SAW resonator in the overcoupled regime, where $Q_i > Q_C$. We measure resonance frequency fluctuations by tracking the amplitude and phase response while driving 14 modes of the SAW resonator simultaneously for $2.6\times10^4$ seconds, approximately 7 hours and 15 minutes. Additionally, we apply a control tone at 2.41 GHz, off-resonant with every SAW mode. This control tone is used as a reference for the signal fluctuations due to other sources in the environment. For details in experimental setup, see Supplemental Materials \cite{supplemental}. The long measurement duration is used in order to capture any TLS telegraphic switching. The rates for such events have been observed to be in the range of 75 $\mu$Hz - 2 mHz \cite{Burnett2019}. We calculate the frequency from the measured reflection coefficient $S_{11}$ using the parameters extracted from the resonance profile in Fig.~\ref{fig_SAWprofile}(b) and Eq.~\ref{S11eqn}.
\begin{figure*}
\centering 
\includegraphics[width = 0.9\linewidth] {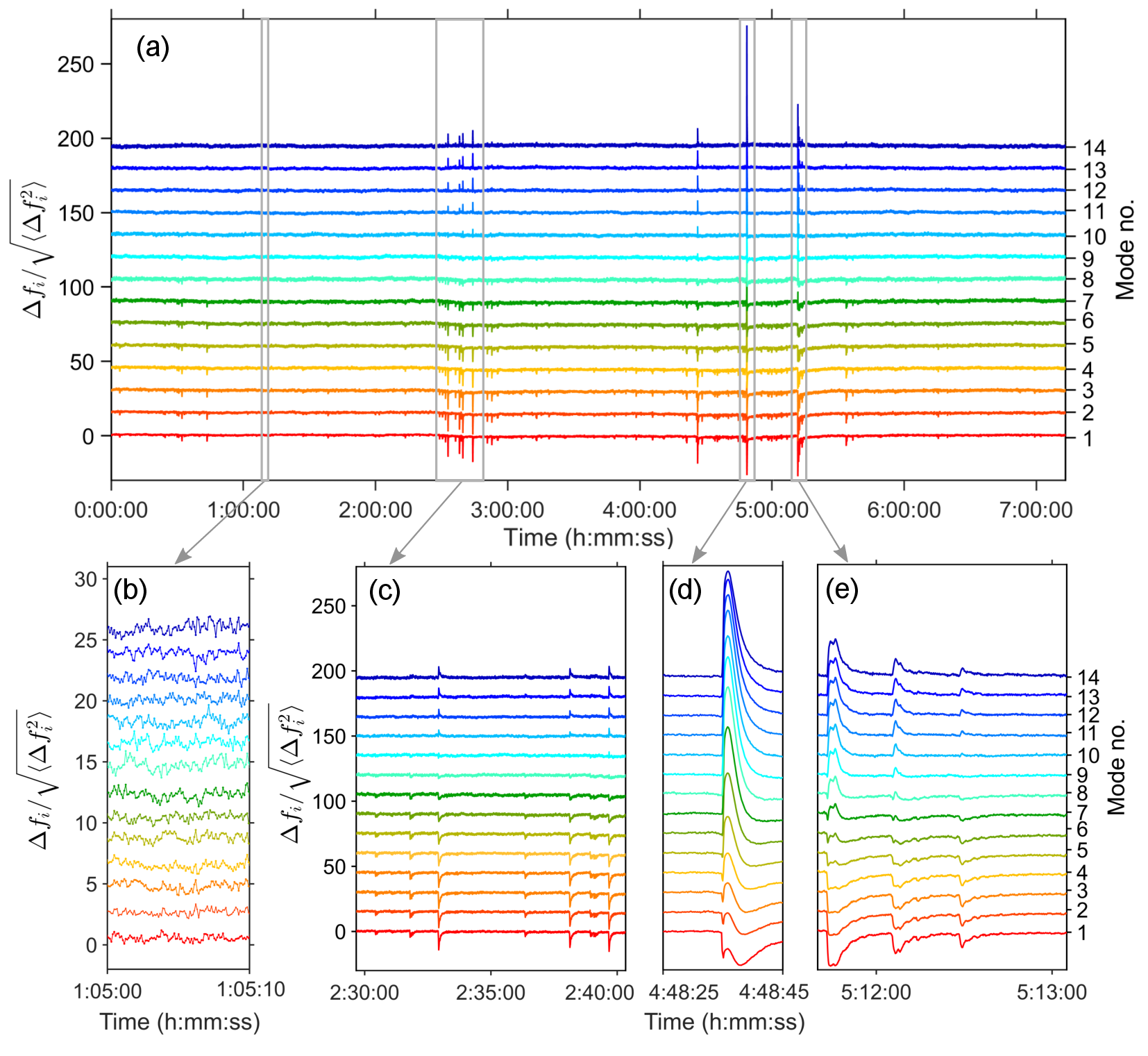}
\caption{(a) Normalized time traces of SAW resonance frequency fluctuations of the resonator mode 1 up to mode 14. All traces are measured simultaneously for approximately 7 hours and 15 minutes. The traces are vertically shifted for better visualization. Bottom panel: Time traces in specific time windows highlighted by grey boxes in the top panel. (b) The resonance frequencies in the ``quiet''  time window where fluctuations are small. Note that the traces are shifted by a different proportion from (a). (c) - (e) The frequency noise in these time windows show additional burst events superimposed with the TLS-induced noise. (See discussion in the main text.)} 
\label{fig_TimeTrace_Freq}
\end{figure*}
\par The time traces in Fig.~\ref{fig_TimeTrace_Freq}(a) show resonance frequency fluctuations of the 14 different modes within the resonator band of 2.37-2.40 GHz. We observe a baseline of frequency fluctuations with uniform amplitude, superimposed with occasional striking large-amplitude fluctuations. There are ``quiet'' time windows where only small fluctuations are present. For instance, in the time window highlighted in Fig.~\ref{fig_TimeTrace_Freq}(b), there are no burst events. The noise amplitude is small, 50 Hz on average, compared to the largest fluctuation of 20 kHz. There is no obvious relationship between fluctuations in different modes. For a device having a small mode volume, we would expect to see a clear telegraphic shift caused by an individual TLS \cite{Niepce2021}. In contrast, our SAW resonator has a large mode volume. Thus, we can observe frequency drifts resulting from many weakly-coupled TLSs, which makes telegraphic features indistinct.
\par We also observe time windows with large fluctuation features. These noise bursts appear randomly in our measurements separated in time by minutes or hours, as shown in Fig.~\ref{fig_TimeTrace_Freq}(c)-(e). In these periods, we recognize similar features in all traces, indicating significant correlations between them. Figure~\ref{fig_TimeTrace_Freq}(c) shows a series of substantial frequency shifts followed by a relaxation tail. The shifts are negative in the low mode numbers. Their magnitude reduces as the mode number increases. Above mode 10, the frequency shift changes sign to positive, while the magnitude grows as the mode number increases further.
\par Figure~\ref{fig_TimeTrace_Freq}(d) presents frequency noise with a large amplitude. At 4:48:35 hours, we observe a frequency increase followed by a relaxation tail. This feature superimposes with a different frequency shift, which has a weaker amplitude and can be either positive or negative. While the burst event shows positive shifts in all modes, this second event has a negative sign in the first nine modes. Above mode 10, both shifts are positive, resulting in anomalously high amplitudes, on the order of a few kilohertz. In Fig.~\ref{fig_TimeTrace_Freq}(e), we can see the characteristic of the noise of the second kind, where the magnitude gradually changes and switches signs. The magnitude of fluctuations decreases with increasing mode number before the frequency shift becomes positive, similar to the features observed in Fig.~\ref{fig_TimeTrace_Freq}(c). 
\par
\begin{figure}
\includegraphics[width=0.8\linewidth] {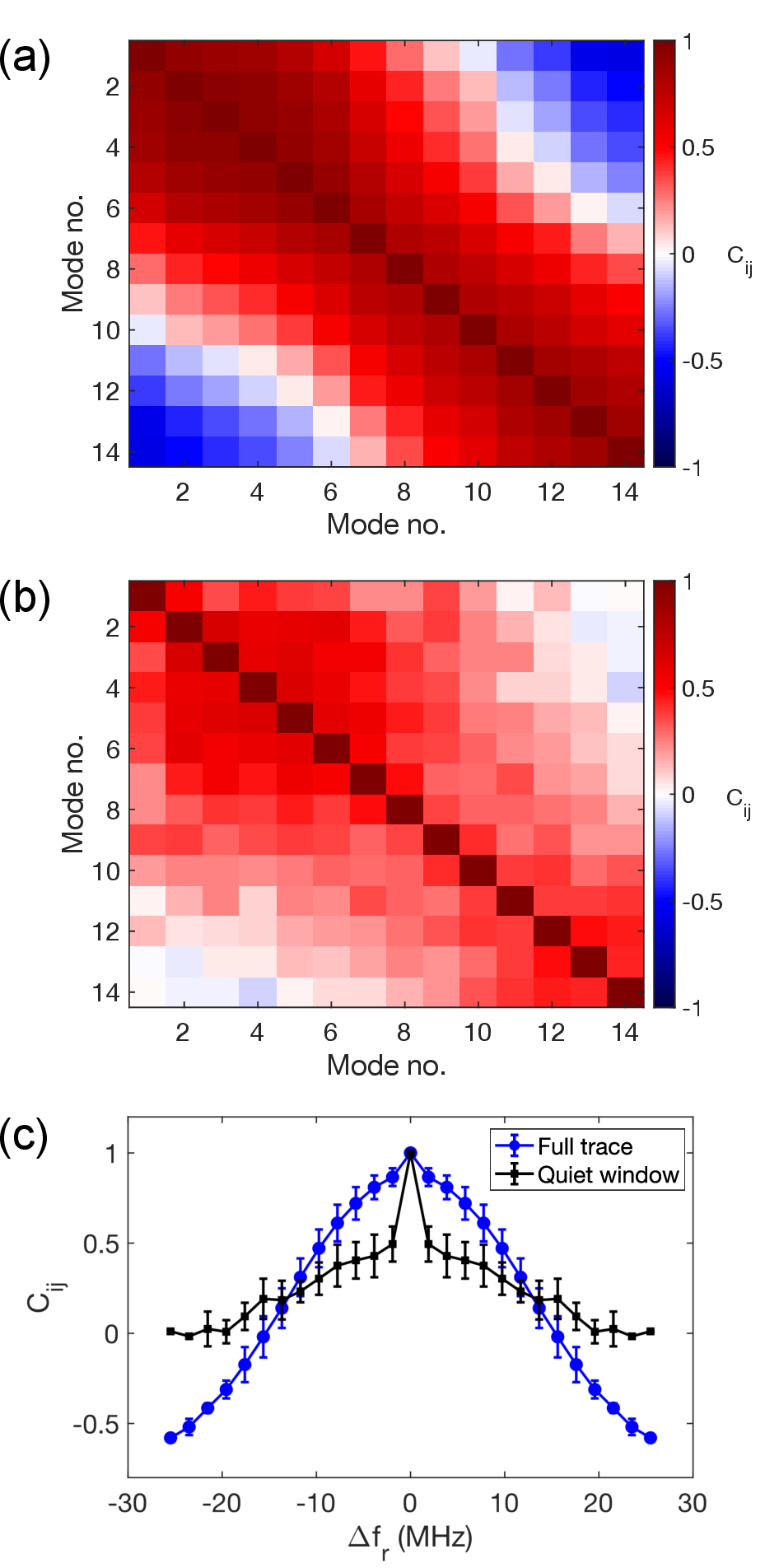}
\caption{(a) The matrix showing the correlation coefficients of the frequency noise measured for 7 hours and 15 minutes, shown in Fig.~\ref{fig_TimeTrace_Freq}(a). Note that the correlation diminishes and becomes negative when the pair of modes is far-detuned. (b) The correlation matrix calculated from the quiet time window shown in Fig.~\ref{fig_TimeTrace_Freq}(b). For this specific period, the correlations between modes still manifest a diminishing trend. However, we no longer observe clear negative correlations between modes. (c) Averaged correlation coefficients as a function of detuning calculated from the whole time trace (blue) and the quiet window (black). For the correlations of the full time trace, the sign of the correlation switches from positive to negative when the detuning is approximately 18 MHz. Note that the plot is symmetric as $C_{ij}=C_{ji}$.}
\label{fig_Corrcoef}
\end{figure}

 To determine the relationship between frequency noise in different modes $i$ and $j$, we compute the correlation coefficients $C_{ij}$, which is given by
\begin{equation}
    C_{ij} = \frac{\langle \Delta f_i \Delta f_j\rangle}{\sqrt{\langle \Delta f_i ^2 \rangle  \langle \Delta f_j ^2 \rangle}},
    \label{Eqn_corrcoef}
\end{equation}
where $\langle\ \rangle$ denotes the expectation value. Figure~\ref{fig_Corrcoef}(a) illustrates the correlation matrix of the frequency noise calculated from the entire measurement (7 hours and 15 minutes). The frequency noise correlation reduces with increasing detuning, i.e., modes closer in frequency show stronger correlations than far-detuned modes.  As suggested by Eq.~\ref{Eqn_JC}, at a given frequency $f_r$, a TLS far from resonance contributes to the frequency noise less than a near-resonant TLS. Furthermore, we obtain negative correlations at a certain detuning, implying that two modes have frequency shifts in opposite directions. This result is not surprising, given that many large frequency shifts shown in Fig.~\ref{fig_TimeTrace_Freq}(c)-(e) explicitly show opposite signs.
\par Based on the STM, the anti-correlated noise bursts are not likely to be induced by TLSs. Under a continuous drive, near-resonant TLSs in the ground state can be excited by the SAW field. If TLSs are saturated, they do not contribute to resonance frequency shifts \cite{Andersson2021}. Thus, saturating TLSs in a specific part of the frequency spectrum will result in an effect similar to the removal of TLSs and their associated dispersive shift, pushing neighboring SAW modes toward the saturated region. The modes that are positively detuned from the saturated TLSs will experience negative frequency shifts, while those that are negatively detuned from the saturated TLSs will have positive shifts. Therefore, we would expect anti-correlated frequency shifts for the modes on opposite sides. Such frequency shifts caused by saturating TLSs are in the reverse direction to the results observed in our measurements. 
\par According to Eq.~\ref{Eqn_JC}, the dispersive interaction between a resonator and a ground-state TLS will shift frequencies away from each other. These frequency shifts will give rise to negative correlation coefficients, which are similar to the effects observed in our experiment. Anti-correlated frequency noise could arise from the sudden appearance or elimination of TLSs at specific frequencies, and such ``scrambling'' of the TLS distribution has been observed with superconducting qubits \cite{Thorbeck2023}. That process, however, does not show the slow relaxation of mode frequencies observed here and is not consistent with stronger anti-correlations between further-detuned modes.

\par Considering the frequency noise only in a quiet period, we do not observe clear negative correlations, as shown in Fig.~\ref{fig_Corrcoef}(b). From the correlation matrices, we can calculate the average correlation as a function of detuning between modes. Figure~\ref{fig_Corrcoef}(c) shows that the correlation coefficients between all pairs of modes show a smooth variation. In the quiet time window, the noise correlation reduces from unity to zero. For the full time trace, the correlation switches its sign when the detuning exceeds 18 MHz. The diminishing positive correlation characteristic agrees with our current understanding of TLS interaction. However, this mechanism cannot explain the anti-correlations during the burst periods.
\par
\begin{figure}
    \centering
    \includegraphics[width=\linewidth]{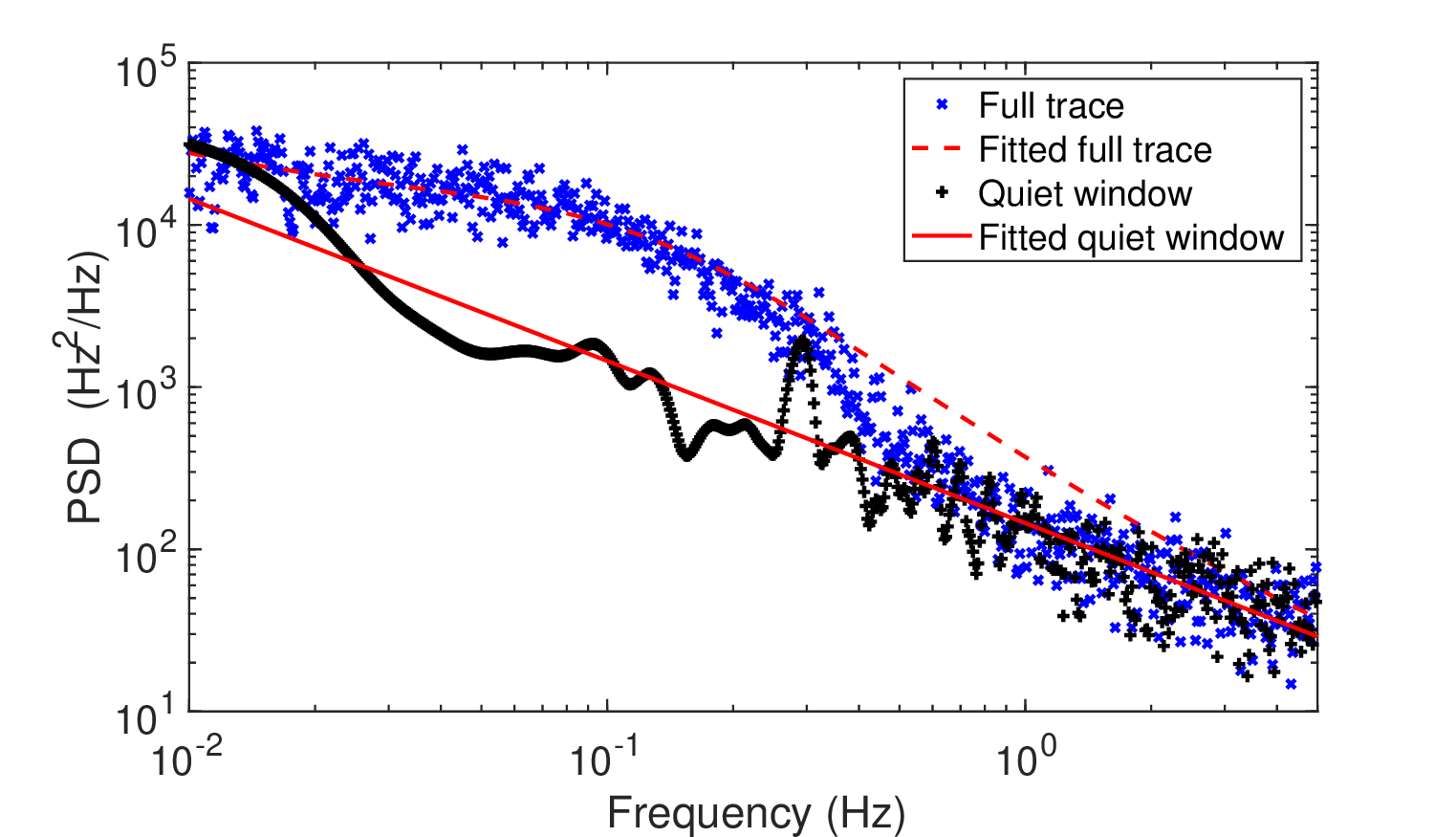}
    \caption{Noise power spectral density of the frequency fluctuations in mode 8 calculated from the entire trace and from the quiet time window only (approximately 400 second-long). The lines represent the fitted power law noise model (Eq.~\ref{Eqn_NoiseModel}) and the sum of the power law noise and the Lorentzian noise model (Eq.~\ref{Eqn_NoiseModel} + Eq.~\ref{Eqn_Lorentzian}).}
    \label{fig_PSD}
\end{figure}

 We further perform frequency domain analysis by calculating the noise power spectral density $S_f$. We consider the standard power law noise, where the noise spectrum is expressed by \cite{Rubiola2008}
\begin{equation}
    S_f(f) =\frac{h_{-1}}{f}+h_0.
    \label{Eqn_NoiseModel}
\end{equation}
Here, $h_{-1}$ and $h_0$ denote the $1/f$ and the white noise level, respectively. The power spectral density (PSD) of the frequency noise in mode 8 is plotted in Fig.~\ref{fig_PSD}. The PSD calculated from the quiet time window exhibits $1/f$ characteristic, which is anticipated for the TLS-induced frequency noise \cite{Burnett2013}. For the full trace, where the frequency bursts are present, the PSD at low frequencies is one order of magnitude higher than that of the quiet window, indicating that the anomalous frequency fluctuations dominate in this frequency range. See Supplemental Material  for time-domain analysis \cite{supplemental}. We fit a sum of $1/f$ and Lorentzian frequency noise  $PSD (f) = S_L(f) + S_f(f)$, where
\begin{equation}
S_L (f) = \frac{4A^2\tau_0}{1+(2\pi f \tau_0)^2}.
    \label{Eqn_Lorentzian}
\end{equation}
Here, $A$ and $\tau_0$ are the Lorentzian noise amplitude and the characteristic time scale, respectively. With this model, we obtain the Lorentzian noise amplitude $A = 53.12$ Hz, the Lorentzian characteristic time $\tau_0 = 1.22$ s, the $1/f$ noise power $h_{-1} = 145.0$ Hz$^2$, whereas the white noise level is negligible. The characteristic time falls in the range of the characteristic time of relaxation tails 1.2-3.0 s, observed in the time traces in Fig.~\ref{fig_TimeTrace_Freq}. This fit does not capture the steep rolloff of the PSD at higher frequencies, indicating a non-Lorentzian spectrum of the slow frequency excursions. Our analysis in both time and frequency domains suggests that the frequency noise cannot be purely induced by TLSs and the large fluctuations are caused by an additional noise source.

\par As shown in Fig.~\ref{fig_TimeTrace_Freq}, frequency noise anomalies have an abrupt change followed by a relaxation tail. This characteristic resembles the effect of ionizing radiation in high-kinetic inductance superconducting devices caused by cosmic rays \cite{Cardani2021}. In that case, the long-time transient effects are due to long recombination times of quasiparticles above the superconducting gap, a mechanism absent in our mechanical resonator. The expected flux of high-energy cosmic muons is around 10 per hour across the chip cross-section \cite{Vepsalainen2020, Cardani2021, Tsuji1998}, a rate consistent with our data, although there is no obvious interaction mechanism. An ionization event induced by radiation could lead to the generation of electron-hole pairs which get trapped in deep-level defects with long relaxation times similar to the timescale of observed fluctuations \cite{Zuleeg1985,Peaker2018}. The correlations of the frequency noise between different modes vary smoothly with detuning and show a uniform behaviour where the sign of frequency excursions are consistent across different events. This suggests a geometrical effect arising from the spatial distribution of a single or small number of defects and the acoustic field variation with mode number. Further investigation is necessary in order to pin down the origin of anti-correlated fluctuations. 

\par In conclusion, we have reported two distinctive noise processes in a surface acoustic wave resonator. There is TLS-induced noise, showing diminishing correlations with increased detuning and a noise spectral density with $1/f$ characteristics. This noise signature provides additional evidence for TLSs as the origin of phase noise in quantum devices. The correlated nature of phase noise suggests that quantum information encoded in different modes of the same resonator undergo reduced relative dephasing. In the presence of TLS noise, a multimode quantum memory may therefore provide improved preservation of entanglement compared to physically separated modes. Additionally, we observe anomalous frequency fluctuations, whose origins are not clear but show some resemblance to the previously observed impact of cosmic radiation. Our multimode probe scheme reveals striking negative correlations in these fluctuations, whose noise power spectral densities are approximately represented by a Lorentzian noise model.

\begin{acknowledgments}
This work was supported by the Knut and Alice Wallenberg foundation through the Wallenberg Center for Quantum Technology (WACQT), and the Swedish Research Council, VR. JHC is supported by the Australian Research Council Centre of Excellence programme through Grant number CE170100026 and the Australian National Computational Infrastructure facility.
\end{acknowledgments}

\section*{Data availability}
The data generated and analyzed in this study are available upon reasonable request.

\section*{Author contributions}
NT performed the measurements and analysis. NT and GA developed the conceptual idea behind the experiment. JHC provided theory support. PD and GA supervised the project. NT wrote the manuscript with input from all coauthors.

\section*{Competing interests}
The authors declare no competing interests.
\bibliography{reference}

\clearpage
\newpage
\onecolumngrid
\renewcommand{\thefigure}{S\arabic{figure}}
\renewcommand{\theequation}{S\arabic{equation}}
\setcounter{figure}{0}
\setcounter{equation}{0}
\centering{\Huge{Supplemental materials}}
\flushleft
\section{Experimental setup}
\begin{figure}[ht]
    \centering
    \includegraphics[width = 0.8\linewidth]{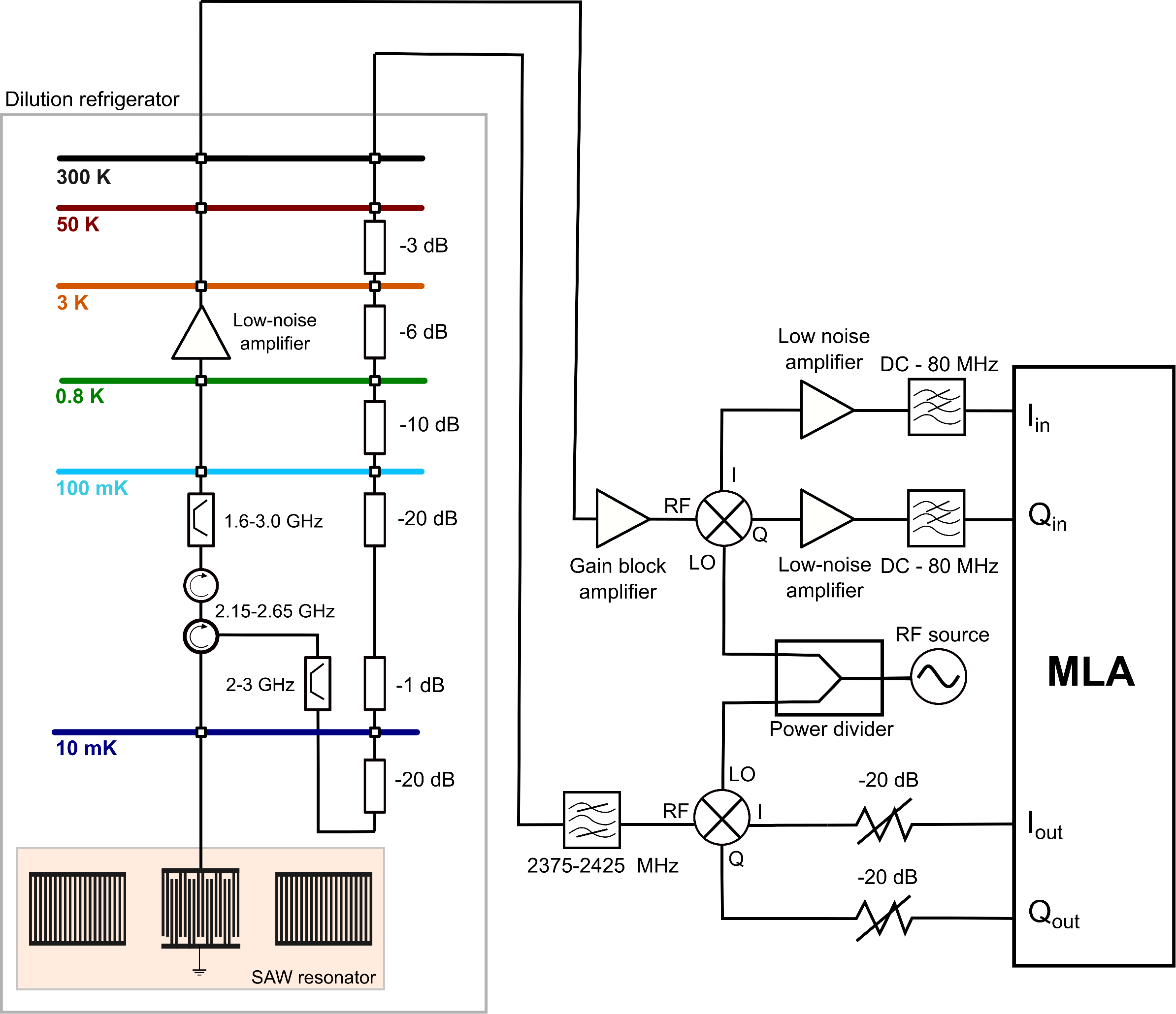}
    \caption{A schematic of our experimental setup shows a SAW resonator installed in a dilution refrigerator and a circuit diagram for measuring the reflection coefficient of the SAW resonator.}
    \label{fig_config}
\end{figure}

The SAW resonator is mounted in a dilution refrigerator and cooled down to around 10 mK. We measure the reflection coefficient of the device using a multi-frequency lock-in amplifier (MLA), which can measure signals up to 32 tones simultaneously. We employ frequency modulation circuits for converting RF signals to the detector operating range. A bandpass filter is inserted after the frequency upconversion to eliminate unwanted lower sidebands. There are 14 modes of SAW available within the 2.37 - 2.40 GHz range. A schematic representation of our experimental setup is given in Fig.~\ref{fig_config}.

\section{Phase fluctuation time traces}
We extract the resonance frequencies from the SAW resonator profile shown in Fig. 2(a). Additionally, we define a control tone at 2.41  GHz, which is outside the SAW bandwidth and located on the positive side relative to all SAW modes. The drive power of the control tone is -47 dBm, equal to SAW drive tones. We measure the SAW resonator profile shown in Fig.1(b) under the presence of the control tone so that it reflects the same condition as in the frequency fluctuation measurement.
\par
\begin{figure*}[ht]
    \centering 
    \includegraphics[width=\linewidth]{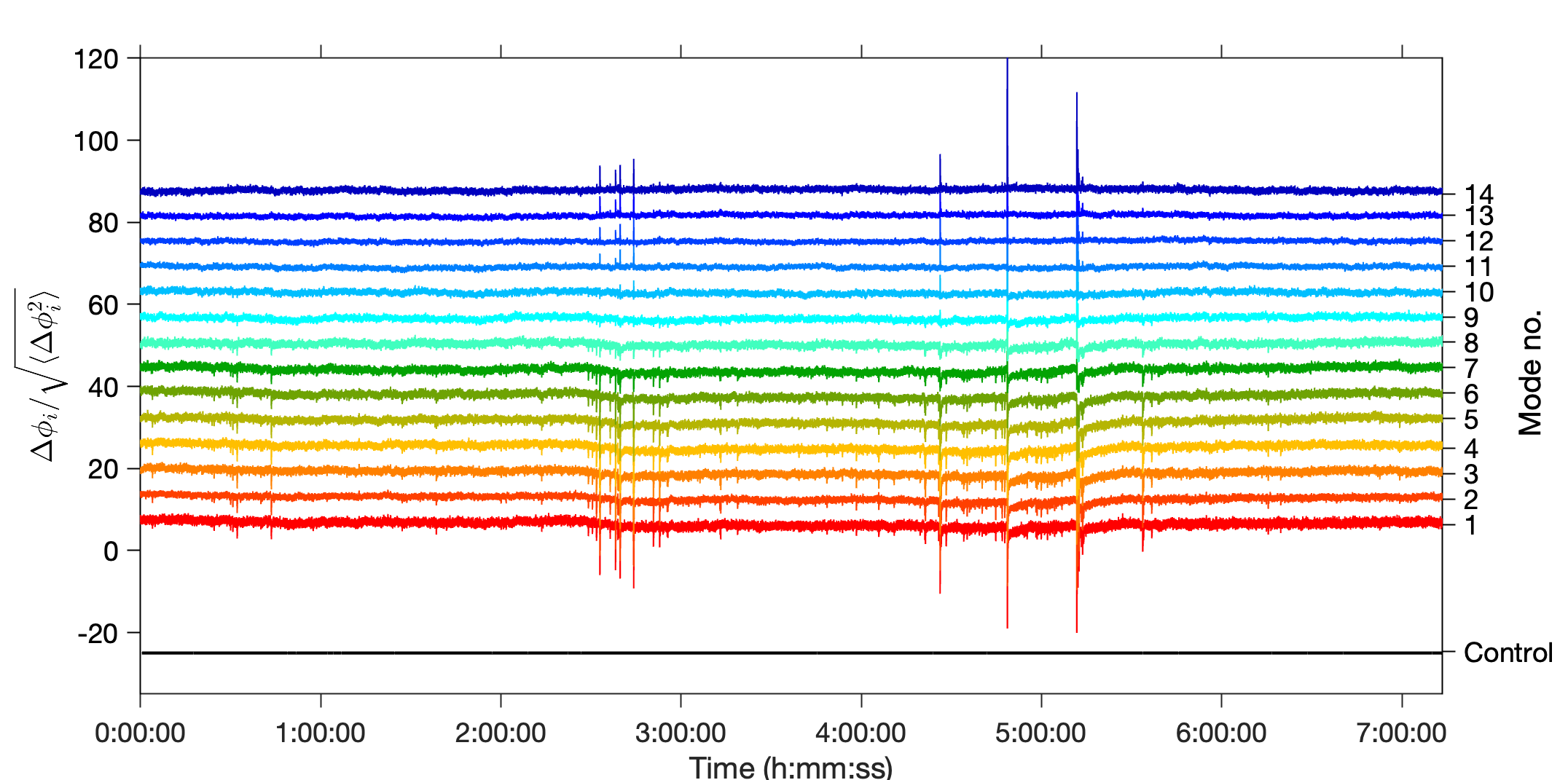}
    \caption{Normalized time traces of phase fluctuations of the control tone and 14 modes of the SAW resonator. All traces are vertically shifted for better visualization. The resonator probing tones show fluctuations, whereas the control tone is rather steady for the entire measurement.}
    \label{fig_phase_timetrace}
\end{figure*}

 We measure reflection coefficients $S_{11}$ of 14 SAW resonance modes and the control tone simultaneously for 7 hours 15 minutes. The control tone serves as a reference for comparing with the SAW probing tones so that we can identify fluctuations from other noise sources in the environment. The SAW probing tones exhibit continuous fluctuations while the control tone shows insubstantial frequency drift, as shown in Fig.~\ref{fig_phase_timetrace}. The amplitude and phase fluctuations of the control tone show no correlation with any other SAW modes. Note that the amplitude and phase of the control tone cannot be converted to frequency as this is a reference tone that does not correspond to any resonance mode.

\section{Time-domain analysis}
We perform frequency domain analysis by estimating the noise power spectral density using Welch's method. However, power spectral density can create more artifacts or uncertainty. Thus, we compute the Allan deviation (ADEV) $\sigma_y$, which is a time-domain analysis tool for analyzing a slow noise process and addressing the source of fluctuation \cite{Rubiola2008}. The mathematical expression of $\sigma_y$ of the $1/f$ and the white noise is written as \cite{Barnes1971}:
\begin{equation}
 \sigma_y(\tau) = \sqrt{2\ln (2) h_{-1}} + \sqrt{\frac{h_0}{2\tau}},
 \label{eqn_ADEV_flicker}
\end{equation}
while ADEV of the Lorentzian noise process is expressed by \cite{Vanvliet1982}:
\begin{equation}
    \sigma_{y_L}(\tau) = \frac{A\tau_0}{\tau}\left( 4e^{-\tau/\tau_0}-e^{-2\tau/\tau_0}+2\frac{\tau}{\tau_0}-3\right)^{1/2}.
    \label{eqn_ADEV_Lorentzian}
\end{equation}
\par
\begin{figure}[ht]
    \centering
    \includegraphics[width = 0.7\linewidth]{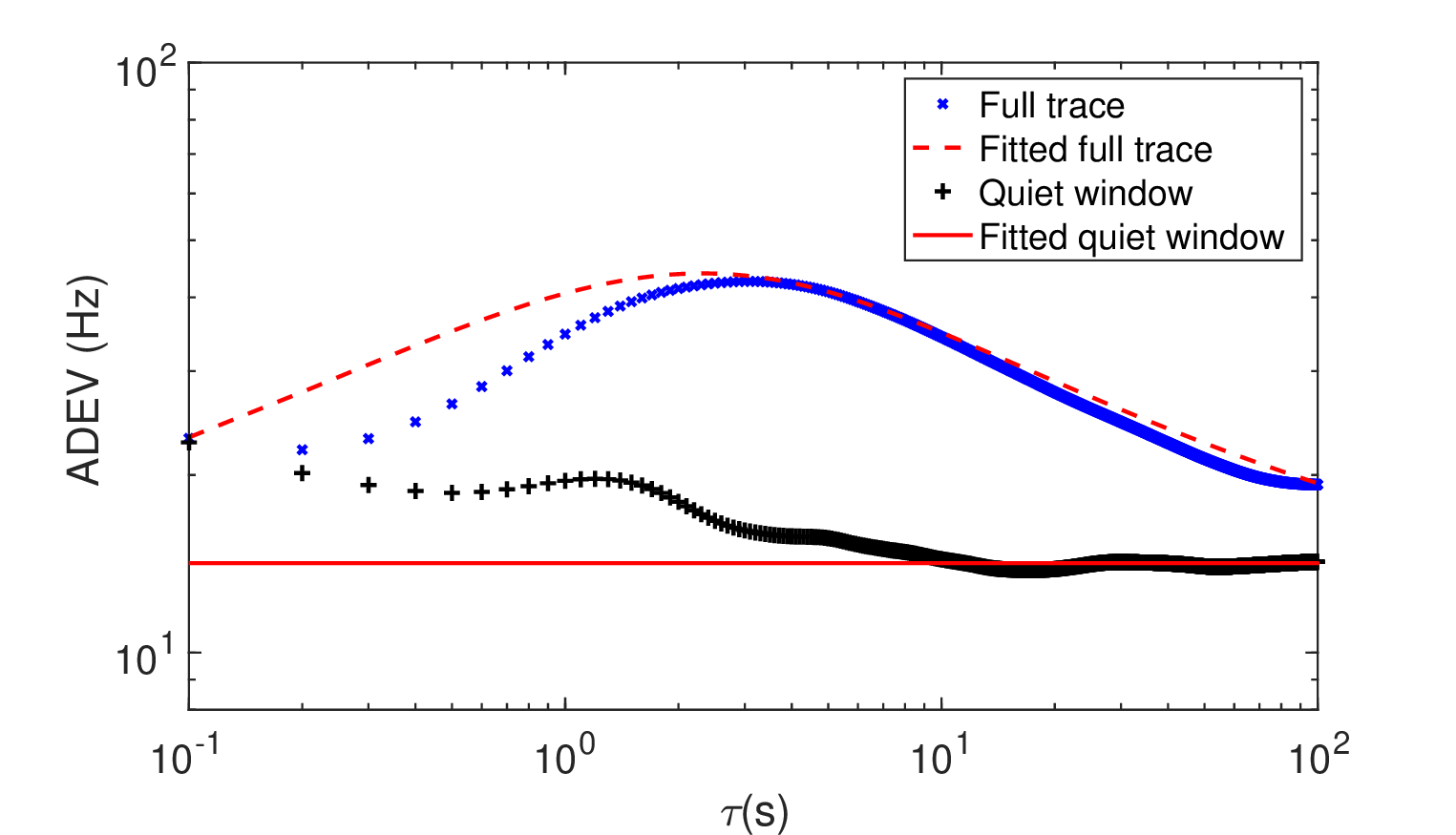}
    \caption{Allan deviation of the resonance frequency mode no.8 calculated from the entire trace and the quiet time window. The lines represent the fitted power law noise model (Eq.~\ref{eqn_ADEV_flicker}) and the sum of the power law noise and the Lorentzian noise model (Eq.~\ref{eqn_ADEV_flicker} +~\ref{eqn_ADEV_Lorentzian}).}
    \label{fig_ADEV}
\end{figure}

 Figure~\ref{fig_ADEV} presents the comparison between the pure $1/f$ characteristics and the combination of the $1/f$ with the Lorentzian noise type. The ADEV of the resonance frequency of the full time trace has a local maximum, which is a characteristic of the Lorentzian noise process, whereas the noise in the quiet window shows a relatively flat ADEV. The parameters obtained from fitting are $A = 53.12$ Hz, $\tau_0 = 1.22$ s, $h_{-1} = 145.0$ Hz$^2$, and $h_0=0$.

\section{Power dependence analysis}

\par
\begin{figure}[ht]
    \centering
    \includegraphics[width = 0.7\linewidth]{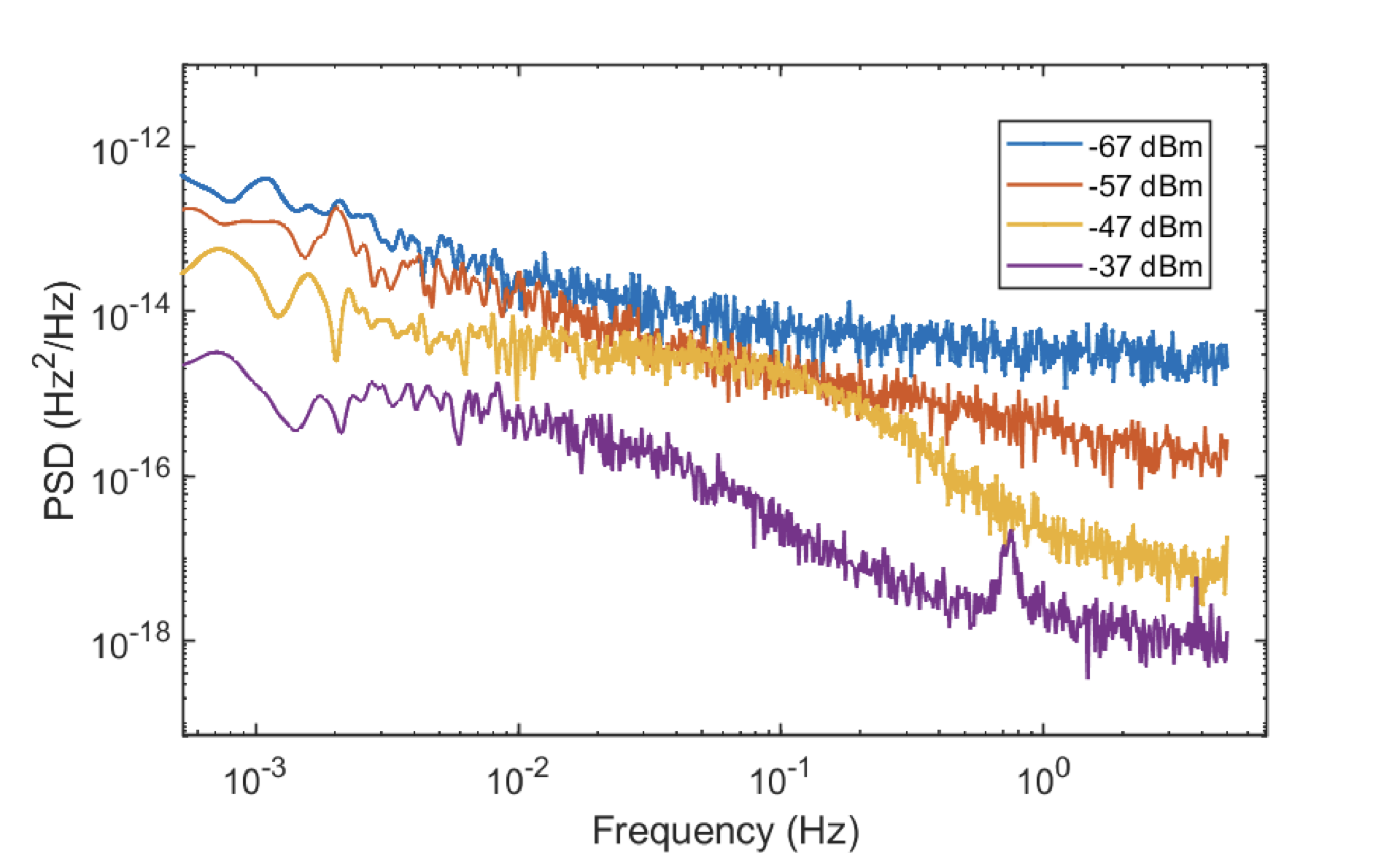}
    \caption{Power spectral density of noise from SAW resonator mode no. 8 obtained using varied drive powers. Generally, the noise level is suppressed with increased SAW driving power, except for drive power of -47 dBm. At this power, we observe anomalous frequency fluctuations exhibiting non-TLS  behaviors.}
    \label{fig_power_dependence}
\end{figure}

 We also study the frequency fluctuations at different probe powers. The noise power spectral density (PSD) at varied drive powers is shown in Fig.~\ref{fig_power_dependence}. The plot shows that the noise level reduces as the drive power increases which is in agreement with the prediction of the STM \cite{Gao2007} and previous findings \cite{Gao2008,Barends2010}.
 
Deviations from the $1/f$ trend are most pronounced for the -47 dBm probe power in the $10^{-4}$ to 1 Hz frequency range, where the Lorentzian noise characteristic is present (See Fig.4 in the main article). At lower power these fluctuations are not visible and PSD follow the theoretical prediction of TLS behavior.

\end{document}